\documentstyle[11pt,aaspp4]{article}

\begin{document}
\title{Production of star-grazing and impacting planetesimals via orbital 
migration of extrasolar planets
}

\author{
A.~C.\ Quillen\altaffilmark{1}   \&
M.\ Holman\altaffilmark{2}   
}
\altaffiltext{1}{Steward Observatory, The University of Arizona, Tucson, AZ 85721;
 aquillen@as.arizona.edu}
\altaffiltext{2}{Harvard-Smithsonian Center for Astrophysics,
60 Garden St., Cambridge MA, 02158; mholman@cfa.harvard.edu}

\begin{abstract}

During orbital migration
of a giant extrasolar planet via ejection of planetesimals 
(Murray et al.~1998), inner mean motion resonances can be strong
enough to cause planetesimals to graze or impact the star.   
We integrate numerically the motions of particles which pass through
the 3:1 or 4:1 mean motion resonances of a migrating
Jupiter mass planet.  
We find that many 
particles can be trapped in the 3:1 or 4:1 resonances and pumped
to high enough eccentricities that they impact the star.
This implies that for a planet migrating a substantial
fraction of its semi major axis, a significant fraction of its mass
in planetesimals could impact the star.  
This process may be capable of enriching the metallicity of the star,
and at a time when the star is no longer fully convective.
Upon close approaches to the star
the surfaces of these planetesimals will be sublimated.
Orbital migration should cause continuing production
of evaporating bodies, suggesting that this process 
should be detectable 
with searches for transient absorption lines in young stars.  
The remainder of the particles will not impact the 
star but can be subsequently ejected by the planet
as it migrates further inwards.
This allows the planet to migrate a substantial fraction
of its initial semi-major axis via ejection of planetesimals.

\end{abstract}


\section {Introduction}

In the standard scenario for solar system formation, solid
material in the disk forms rocky or icy bodies called planetesimals.
These then accumulate in certain regions to form planets.
The moderate detection rate of dusty disks
with IRAS and ISO in the far infrared, particularly 
surrounding younger stars 
(\cite{aumann}, \cite{becklin}, \cite{beckwith}),
suggest that planet formation is often accompanied by the formation of
belts (e.g., the Kuiper belt and possibly the Main asteroid belt).
Recently spectral features of crystalline silicate material similar to those
observed in comets have also been detected in these disks, suggesting that 
there is asteroidal and cometary material in these disks
(\cite{malfait}, \cite{waelkens}, \cite{pantin}).
The detection of planets orbiting nearby  solar-type stars (e.g., \cite{mayor})
and dusty disks surrounding some of these stars (e.g., \cite{trilling2})
confirms the connection between rocky disk material and planets.
Notably stars with known extra-solar planets have enhanced
metallicities (\cite{gonzalez99_3}; \cite{gonzalez}), 
establishing an as yet unexplained link
between planet formation and stellar metallicities.

The small orbital semi-major axes of many of the newly discovered 
extrasolar planets 
($a< 0.1$ AU) is surprising.  This has resulted in the proposal of two 
classes of planetary orbital migration mechanisms.  
One mechanism involves the transfer of 
angular momentum between a planet and a gaseous disk (e.g. \cite{trilling}; 
\cite{lin}).  The other focuses on resonant interactions between planetesimals
and the planet and the resulting ejection of the planetesimals 
(in extrasolar systems \cite{murray}, and in our solar
system \cite{fernandez} and \cite{malhotra}). 
The first mechanisms suffers
from a fine tuning problem where only a small range of planet and disk masses
would allow migration but not the destruction of planet by the star
(\cite{trilling}).
Metals from planets accreted by the star
could account for the enhanced metallicities of the more massive 
stars with known planets.
However because stars with masses comparable to the sun have 
convective envelopes 
for nearly the the entire time interval over which planets are expected 
to be accreted, encorporation of giant planets
into the star should not be able to enhance the stars metallicity
substantially (\cite{adams}).

The second mechanism involving ejection of planetesimals (\cite{murray}) 
has some advantages over the first mechanism.
Planetesimals affected by the inner resonances can be driven
to extremely high eccentricities and so can impact the star
(\cite{beust}; \cite{gladman}; \cite{moons}; \cite{wisdom}; \cite{farinella}; 
\cite{migliorini};
\cite{ferraz}).  
This would happen at a later time ($\gtrsim 10^7$ years; \cite{murray})
than appropriate for the migration scenario involving a gaseous disk 
($\sim 10^6$ years).  Thus addition of rocky or metallic material will happen
when the stellar convective envelope is small
so that the metals will remain trapped in
the convection zone, rather than mixing into the star.
In this way orbital migration via ejection of planetesimals would more 
naturally explain the enhanced metallicities of stars with massive planets.
As pointed out by \cite{gonzalez_} adding 20~$M_\oplus$ (earth masses) of
asteroidal material to the convection zone of the star is sufficient to increase 
the enhanced metallicities of a solar type star by 
$ \Delta {\rm [Fe/H]} \sim 0.1$ dex.
For a planet to migrate a significant fraction
of its initial semi-major axis 
roughly {\it its} mass of planetesimals must be ejected from the system
(\cite{murray}).  Since in the inner solar system 
this material is expected to be asteroidal or rocky
this could result in a significant fraction of a Jupiter
mass planet ($M_J = 310 M_\oplus$) impacting and 
becoming incorporated into the star.

In this paper we concentrate on the mechanism
for producing star grazing planetesimals explored
by \cite{beust_} to account for the transient
absorption lines observed against beta Pictoris 
(e.g., \cite{crawford}; \cite{lagrange96}).  
In this context a star grazing planetesimal
approaches within 10 stellar radii of the star.  
Mean motion resonances 
(such as the 3:1 and 4:1) with one large moderately eccentric planet, 
can pump eccentricities to 1.0.
In \S 2 using averaged Hamiltonians we plot the range of planetesimal
and planet eccentricities
needed for a given resonance to produce a star impacting
body.   However the region of phase phase that results
in extremely high eccentricity orbits is not necessarily large
since many particles with semi-major axes containing the resonance
will not librate or will not librate to high eccentricities
(e.g., as shown in the contour plots of \cite{yoshikawa90}; 
\cite{yoshikawa91}).  
As a planet migrates we expect many particles
to have orbital elements such that they will not be caught in the active
(or high eccentricity) part of the resonance.
So in \S 3 we estimate via numerical integration
the efficiency of these resonances to produce extremely
high eccentricity particles.  For a series of integrations
we tabulate the numbers of particles which impact the star 
and those which eventually cross the Hill sphere of the planet and are ejected
to large semi-major axes.

\section{When can star impacting planetesimals be produced?}

During the migration of a major planet, mean-motion resonances
will be swept through the disk of planetesimals.
Though secular resonances are also capable driving
particles to extremely high eccentricities  (\cite{levisonb})
they may not necessarily be swept through the disk. 
We also cannot necessarily assume that secular
resonances are always strong in extra-solar systems (\cite{beust})
particularly as the innermost planet becomes more distant from
its neighboring planets.
So we concentrate here on mean motion resonances with one
major planet.

The maximum eccentricity reached by particles librating
in a resonance is extremely sensitive to the eccentricity
of the planet (\cite{beust}; \cite{yoshikawa90}; \cite{moons}).
We expand on the work of \cite{beust_} to determine
what range of eccentricities for a planet are required
to pump particle eccentricities to 1.  We created contour plots 
numerically from the Hamiltonian averaged over time
(as in \cite{beust} and \cite{yoshikawa90}).
For each resonance we then determined what minimum initial 
particle eccentricity is needed for a particle 
to later reach the star ($e=1$).  We estimated
this minimum eccentricity (shown in Fig.~1) for a range of
planet eccentricities, $e_p$.
These contour plots are only extremely weakly dependent on the planet
mass.   We see in Fig.~1 that past a planet eccentricity of 0.3
the 3:1, 4:1, 5:1, 5:2 and 7:2 resonances are all capable
of driving low eccentricity particles to extremely high 
eccentricities.  The eccentricities of the extrasolar planets 
are not restricted to extremely low values (\cite{marcy}).  
This implies that resonances which are capable of
causing star grazing or impacting planetesimals are likely
to exist in almost all of these systems.

\section{Simulation of particles in mean motion resonances 
during orbital migration}

To estimate the efficiency of production of high eccentricity
orbits we numerically integrate the orbits of particles
(using a conventional Burlisch-Stoer numerical scheme) 
during the slow migration of a major planet.
All particles are massless except for the star and
one planet with an eccentricity $\epsilon_p$,
which remains constant throughout the integration.
During the integration we
force the semi-major axis of the planet to drift inwards
at a rate given by the dimensionless parameter 
\begin{equation}
D_a = {da \over dt } {P \over a}
\end{equation}
for $P$ the period of the planet and $a$ its semi-major axis.
$D_a$ is fixed during the integration resulting in 
${da\over dt} \propto  \sqrt{a}$.
Particles are placed in the plane of the planet's orbit  
just within (a few resonance widths) 
the semi-major axis of either the 3:1 or 4:1 resonance.
For each particle the angle of perihelion 
and the mean anomalies were chosen randomly.
Massless particles were integrated until they 
were driven to high eccentricity  ($\epsilon > 0.995$)
and so impact the star,  
or crossed the Hill sphere radius of the planet and were
ejected to semi-major axes larger than the planet.
This took between a few times $10^5$ to $10^6$ periods
measured in units of the initial orbital period of the planet.
In Table 1 we note the initial conditions, migration rates, planet
masses and eccentricities (which remain fixed during the simulation), 
and final particle fates for a set of 10 particle integrations.
In Table 2 we note the resonances operating on the particles
in each simulation prior to impact or ejection.

A sample plot showing eccentricity and semi-major axes for
a run (denoted N8) are shown in Fig.~2.
Almost at all times particles are strongly affected by resonances.  
When a particle crosses the 3:1 or 4:1 resonance it may be 
trapped in a high eccentricity
region of the resonance.  Then the particle can be pumped to extremely
high eccentricities and impact the star.   We find that both the 3:1 and 4:1
resonances cause impacts.  However if the particle does not
remain trapped in the resonance it can later on be caught in
another resonance.   For example, we observe that particles not removed by 
the 3:1 may later on be caught in the 5:2 or 7:3 resonances and particles 
not initially affected by the 4:1 may subsequently be caught in the 
3:1, 7:2 or 8:3 resonances (see Table 2).
If the particle is trapped or strongly affected by a resonance
nearer to the planet (such as the 8:3 resonance) 
then it has a higher chance of being ejected than hitting the star.
In the slower migration rate simulations (N5, M5) we see that even 
minor resonances
such as the 11:3, 10:3, 11:4 cause jumps in the semi-major axis as
the particle crosses the resonance.  However only the 3:1 and 4:1 
are strong enough (and with large enough regions in phase space)
that particles are trapped in them for long periods of time.
These resonances are responsable for the majority of impacts.

In Fig.~2a we see that particles trapped in the 3:1 and 4:1 resonances can 
make multiple close approaches to the star.
During a close approach the surface of a planetesimal will graze
the star and so be sublimated it.
Thus we would predict that a migrating planet would cause
continuing production 
of `falling evaporative bodies',  as proposed to explain the transient
absorption lines observed against beta Pictoris and other stars
(e.g., \cite{beust}; \cite{crawford}; \cite{lagrange96}; \cite{grady}).
We see in our simulations that 
more than one resonance can cause star grazers.
If star grazers are produced by more
than one resonance then particles could approach
the star from different angles with respect to the planet's angle of 
perihelion.  This might provide an alternative explanation for
the occasional blue-shifted event on beta Pictoris (\cite{blueshift}).


Even though the 3:1 and 4:1 resonances can pump
eccentricities to 1.0, in every simulation (see Tab.~1) we find 
particles which pass
through these resonances that are not pumped
to high eccentricities and so removed from the system
by an impact with the star.
These particles can later be ejected by the planet.
While the 3:1 and 4:1 resonances can reduce the surface density
in a disk of planetesimals, they do not create a hole as they
are swept through the disk.  
If the density of planetesimals is
high enough, a planet migrating as a result of
ejection of planetesimals can migrate to within
its original (at formation) 3:1 or 4:1 mean motion resonances.
This would allow a planet to migrate a substantial
fraction of the planet's semi-major axis via ejection of 
planetesimals.  

Particles which impact the star loose all their angular momentum
to the planet which would reduce the planet's
eccentricity.  During the time they are trapped in a resonance
their semi-major axes decreases.  This implies
that the planet will gain energy.
However the decrease in semi-major axis was typically less than $30\%$
of the particle's initial semi-major axes,   
so the energy gained by the planet from trapped particles
should be small compared to that lost from ejected particles.

We did not find that the fraction of impacts was strongly dependent on 
the planet migration rate,  
initial particle conditions, or planet eccentricity.  
However more particles should be integrated to verify this.
We would have expected that slower migration rates, 
more massive planets, lower eccentricity initial
particle eccentricities, and higher planet eccentricities 
would result in an 
increase in the efficiency of trapping particles in resonances
and so in producing impacts.  However the number of resonances
operating on each particle makes it difficult to predict the final
states.  For example when the migration rate
is fast or the planet eccentricity is lowered then  we found that
weaker resonances such as the 4:1 or 7:2 did not affect the particles much,
however the 3:1 was still strong enough to cause impacts.  

\subsection{Survival until impact}

In our integrations we can estimate  
the timescale for the eccentricity to reach $\gtrsim 0.995$. 
While some particles impact the star
on a very short timescale (e.g., particle 1 in Fig.~2), others
are slowly pumped to high eccentricities (e.g., particles
5,6,7,and 8 in Fig.~2).
For the slower approaches the particle or planetesimal
could make $\sim 10^4 - 10^5$ close passages to the star before impact.
When the migration rate was slower 
particles typically experienced larger numbers of close passages
before impact.  We have estimated the mass loss from
a rocky body during a free fall time at solar
radius from the sun to be $\sim 30$ cm.
If the planetesimal makes $10^4$ of such close passages then 
a km body will be completely evaporated by a solar type star.  
For particles making multiple close approaches
only large bodies $\gtrsim 1$km will survive until
impact.  Notably the size distribution of
asteroids subsequent to Gyr timescale collisional evolution 
(\cite{davis}; \cite{greenberg})
is expected to be such that most of the integrated disk {\it mass} is contained
in the largest bodies.  
When migration is relatively quick, 
this mechanism could be a way to increase the metallicity of the star, 
despite the fact that the lower mass bodies may not survive until impact.
Smaller bodies which will completely evaporate could
manifest themselves as transient absorption
features, a phenomenon which is observed on beta Pictoris 
and other stars (e.g., \cite{crawford}; \cite{lagrange96}; 
\cite{grady}).  


We now consider whether large bodies are likely to fragment
upon close approach.  If the object is strengthless then 
it is likely to fragment at periapse
only if the density of the object is lower
than the mean density of the star 
(e.g., \cite{asphaug}; \cite{sridhar}).
The mean density of the sun is $\rho \sim 1.4$ g cm$^{-3}$
so that on a solar type star all but the least dense asteroids 
should not fragment and so should survive
until impact.  On lower mass main sequence stars (which are denser), 
however, denser objects
could be fragmented during close passages subsequent to impact.  
On higher mass stars, such as beta Pictoris, even cometary material 
will not be fragmented by the star during close passages.  

\section{Summary and Discussion}

We have presented a series of numerical integrations
of particles initially at low eccentricities which
pass through mean motion resonances with
a major moderate eccentricity migrating planet.
We confirm that the 3:1 and 4:1 resonances can pump
the particles eccentricities to 1.0 and so can cause particles
trapped in them to impact the star or be evaporated by it. 
As a planet migrates through a disk of planetesimals we would
expect continuing production of bodies undergoing close
approaches to the star.
This provides us with a possible observational test.
A recent study finds that beta Pictoris may be quite young ($2\times 10^7$ years;
\cite{agebetapic}).  If orbital migration occurs commonly
during this timescale then a multi-object (or multi-fiber) survey
in young clusters should detect
transient absorption features due to evaporating bodies 
similar to those seen in beta Pictoris and other stars.

Our integrations show that many particles which pass
through these resonances will not be pumped
to high eccentricities and so removed from the system
by evaporation or by impact with the star.
These particles can subsequently be ejected by the planet.
This implies that a planet can migrate 
a significant fraction of its initial semi-major axis via
ejection of planetesimals.

For the faster migration rates,
we estimate that $\gtrsim 1$ km sized rocky bodies will survive
heating from a solar type star during multiple close passages
and so can become incorporated into the convection
zone of the star.  Because we expect that most of the mass 
will be in the most massive bodies,
this migration process may be capable of increasing the 
metallicity of the star.
Planet migration should occur on a $10^7$ year timescale 
(\cite{murray}) so we do not expect
the star to be fully convective during migration.
Metals dumped into the star should remain
in the convection zone of the star.
This scenario therefore offers a plausible explanation
for the metallicity enhancements observed in stars
with extrasolar planets (\cite{gonzalez99}).

To migrate a significant fraction of its semi-major axis
the planet must eject on the order of its mass (\cite{murray}) in 
planetesimals.  If the material ejected is rocky then the original 
proto-stellar disk would have had $\gtrsim 30$ times this mass in
gas and volatiles.  It is not inconceivable that this amount of material
was left in and interior to a Jupiter mass planet after formation.
However planetesimals exterior to the planet forced to 
high eccentricity by a secondary planet may also
be ejected by a planet and so cause its
migration.  Some fraction of these particles will also impact the star
(e.g., as seen in simulations of short period comets, \cite{levison}). 
This suggests another possible link between star grazers and 
impactors and orbital migration. 

\acknowledgments

This work could not have been carried out without
suggestions, discussions and correspondence 
from N.~Murray, C.~Pilachowski, R.~Strom, M.~Sykes, D.~Davis, R.~Greenberg,  
J.~Raymond,
D.~Trilling, D.~Garnett, B.~Livingston, R.~Kudriski, and J.~Lunine.
We also thank G.~Rivlis and K.~Ennico for
donations of computer time and support.
We acknowledge support from 
NASA project numbers NAG-53359 and NAG-54667. 

\clearpage


%


\begin{deluxetable}{lccrclrcccccc}
\footnotesize
\tablecaption{Numerical Integrations}
\tablehead{
\multicolumn{1}{l}{Run} &
\colhead{$\epsilon(t=0)$}                  &
\colhead{$\epsilon_p$}                     &    
\colhead{$M_p/M_*$}                        &          
\colhead{Resonance}                        &          
\colhead{$da$}                             &
\colhead{$D_a$}                            & 
\colhead{$N_{imp}$}                        &        
\colhead{$N_{ej}$}                         \cr      
\multicolumn{1}{l}{(1)} &
\colhead{(2)}                              &
\colhead{(3)}                              &
\colhead{(4)}                              &
\colhead{(5)}                              &
\colhead{(6)}                              &
\colhead{(7)}                              &
\colhead{(8)}                              &
\colhead{(9)}                              
} 
\startdata
  M1  & 0.1 & 0.3 & $10^{-3}$ & 3:1 & 0.03  & $10^{-6}$          &  3  &  7   \cr
  M5  & 0.1 & 0.3 & $10^{-3}$ & 3:1 & 0.015 & $3 \times 10^{-7}$ &  6  &  4   \cr
  M8  & 0.1 & 0.3 & $10^{-3}$ & 3:1 & 0.03  & $3 \times 10^{-6}$ &  3  &  7   \cr
  M7  & 0.3 & 0.1 & $10^{-3}$ & 3:1 & 0.03  & $         10^{-6}$ &  5  &  5   \cr
  M9  & 0.1 & 0.3 & $3\times 10^{-3}$ & 3:1 & 0.03   & $10^{-6}$ &  6  &  4   \cr
      &     &     &           &     &       &                    &     &      \cr
  N1  & 0.1 & 0.3 & $10^{-3}$ & 4:1 & 0.008 & $10^{-6}$          &  8  &  2   \cr
  N5  & 0.1 & 0.3 & $10^{-3}$ & 4:1 & 0.004 & $3 \times 10^{-7}$ &  8  &  2   \cr
  N8  & 0.1 & 0.3 & $10^{-3}$ & 4:1 & 0.02  & $3 \times 10^{-6}$ &  6  &  4   \cr
  N7  & 0.1 & 0.1 & $10^{-3}$ & 4:1 & 0.008 & $10^{-6}$          &  9  &  1   \cr
  N9  & 0.1 & 0.3 & $3\times 10^{-3}$ & 4:1 & 0.008 & $10^{-6}$  &  5  &  5   \cr
  N10 & 0.3 & 0.3 & $10^{-3}$ & 4:1 & 0.02  & $3 \times 10^{-6}$ &  4  &  6   \cr
  N11 &0.05 & 0.3 & $10^{-3}$ & 4:1 & 0.02  & $3 \times 10^{-6}$ &  7  &  3   \cr
  N12 &0.05 & 0.3 & $10^{-3}$ & 4:1 & 0.02  & $         10^{-6}$ &  7  &  3   \cr
\enddata
\tablenotetext{}{
Columns:
(1) Run number;
(2) Initial eccentricity of particles;
(3) Eccentricity of the planet;
(4) Ratio of the planet mass to the stellar mass;
(5) Particles were placed just within this mean motion resonance;
(6) Distance that particles were placed from the initial
location of the resonance in units of the initial planet
semi-major axis.
(7) Dimensionless orbital migration rate (see text);
(8) Number of particles eventually impacting the star ($N_{imp}$)
out of 10 particles integrated;
(9) Number of particles eventually ejected by the planet ($N_{ej}$)
out of 10 particles integrated.
}
\end{deluxetable}


\begin{deluxetable}{lcccccccccccc}
\footnotesize
\tablecaption{Resonances operating prior to impact or ejection}
\tablehead{
\multicolumn{1}{r}{~~~~~ Particle} &
\colhead{0}            &
\colhead{1}            &
\colhead{2}            &
\colhead{3}            &
\colhead{4}            &
\colhead{5}            &
\colhead{6}            &
\colhead{7}            &
\colhead{8}            &
\colhead{9}            \cr
\multicolumn{1}{l}{Run} &
\colhead{}             &
\colhead{}             &
\colhead{}             &
\colhead{}             &
\colhead{}             &
\colhead{}             &
\colhead{}             &
\colhead{}             &
\colhead{}             &
\colhead{}             
}
\startdata
M1& E&   E&   I&   I&   E&   E&   E&   E&   E&   I           \cr
  & 5:2 &5:2 &3:1 &3:1 &5:2 &5:2 &7:3 &7:3 &5:2 &3:1   \cr
M5& E&   E&   I&   I&   I&   E&   I&   I&   E&   I     \cr
  & 7:3 &8:3 &3:1 &3:1 &3:1 &11:4&13:5?&3:1 &8:3 &3:1   \cr
M7& E&   E&   I&   I&   E&   E&   E&   I&   E&   E     \cr
  & 2:1 &2:1 &3:1 &2:1 &5:2 &7:4 &7:3 &2:1 &2:1 &9:4   \cr
M8& E&   E&   I&   I&   E&   I&   I&   E&   E&   I     \cr
  & 5:2 &8:3?&3:1 &3:1 &5:2 &8:3 &3:1 &3:1 &5:2?&3:1   \cr
M9& I&   I&   I&   I&   E&   I&   E&   E&   E&   I     \cr
  & 8:3 &5:2 &3:1 &3:1 &3:1 &3:1 &5:2 &3:1 &?   &3:1   \cr
%
%
N1& I&   E&   I&   E&   I&   I&   I&   I&   I&   I      \cr
  & 7:2 &8:3 &3:1 &8:3 &10:3?&4:1 &3:1 &7:2 &4:1 &10:3?  \cr
N5& I&   E&   I&   I&   I&   I&   I&   I&   I&   E       \cr
  & 3:1 &7:3 &5:2 &4:1 &10:3&4:1 &7:2 &4:1 &3:1 &4:1   \cr
N7& I&   I&   E&   I&   I&   I&   I&   I&   I&   I      \cr
  & 2:1& 3:1 &5:2 &3:1 &3:1 &3:1 &3:1 &3:1 &3:1 &2:1     \cr
N8 &E&   I&   E&   E&   I&   I&   I&   I&   I&   E       \cr
   &8:3 &10:3&8:3 &8:3 &3:1 &4:1 &3:1 &3:1 &4:1 &8:3   \cr
N9 &I&   I&   E&   I&   E&   E&   E&   E&   I&   I     \cr
   &4:1 &7:2 &7:2 &4:1 &3:1 &3:1 &3:1 &3:1 &4:1 &4:1   \cr
N10&E&   I&   I&   E&   I&   E&   E&   I&   E&   E     \cr
  &10:3?&7:2 &10:3?&8:3?&10:3?&8:3?&10:3?&7:2 &10:3?&5:2?  \cr
N11&E&   I&   I&   I&   I&   I&   E&   E&   I&   I     \cr
   &10:3&3:1&10:3?&3:1 &8:3 &5:2 &8:3 &8:3 &8:3 &3:1   \cr
N12&I&   E&   I&   I&   I&   E&   E&   I&   I&   I     \cr
   &4:1 &10:3&7:2 &3:1 &7:2 &7:2 &8:3 &3:1 &3:1 &4:1   \cr
\enddata
\tablenotetext{}{
For each simulation (labeled on the left) 
the final state of each of 10 particles is listed.    
E  refers to ejection by the planet and I refers
to an impact with the star.
Below this is listed the suspected resonance affecting the particle 
prior to ejection or impact.
}
\end{deluxetable}
\vfill\eject

\begin{figure*}
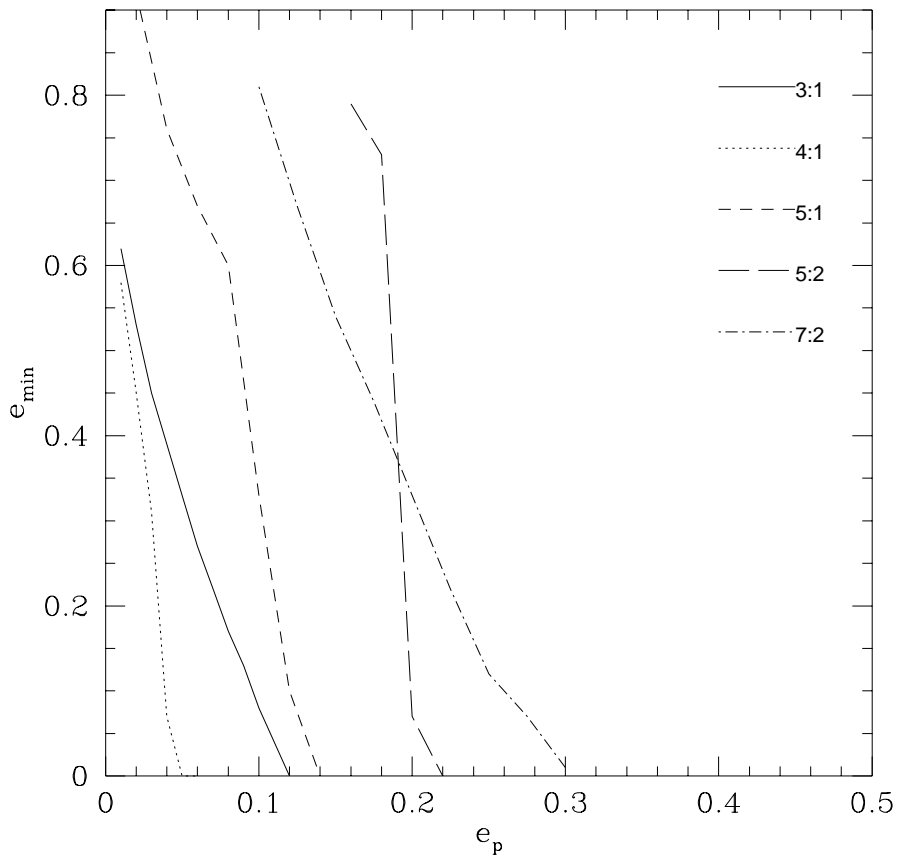

\caption[junk]{
Minimum eccentricity ($e_{min}$) that can be pumped
to a star impacting orbit ($e=1$) for a range of planet
eccentricities ($e_p$).  Each line corresponds to a different
mean motion resonance.
Orbits that become Jupiter crossing (or cross the Hill
sphere radius) are more likely to be ejected from the
system rather than impact the sun.  So we restrict
ourselves to resonances with semi major axis
small enough that high eccentricities can result in stellar
impacts rather than a crossing of the Hill sphere.
The eccentricities of the extrasolar planets
are not restricted to extremely low values (Marcy 1999).
This implies that resonances which are capable of
causing star grazing or impacting planetesimals are likely
to exist in almost all of these extrasolar planetary systems.
\label{fig:fig1} }
\end{figure*}
\vfill\eject

\begin{figure*}
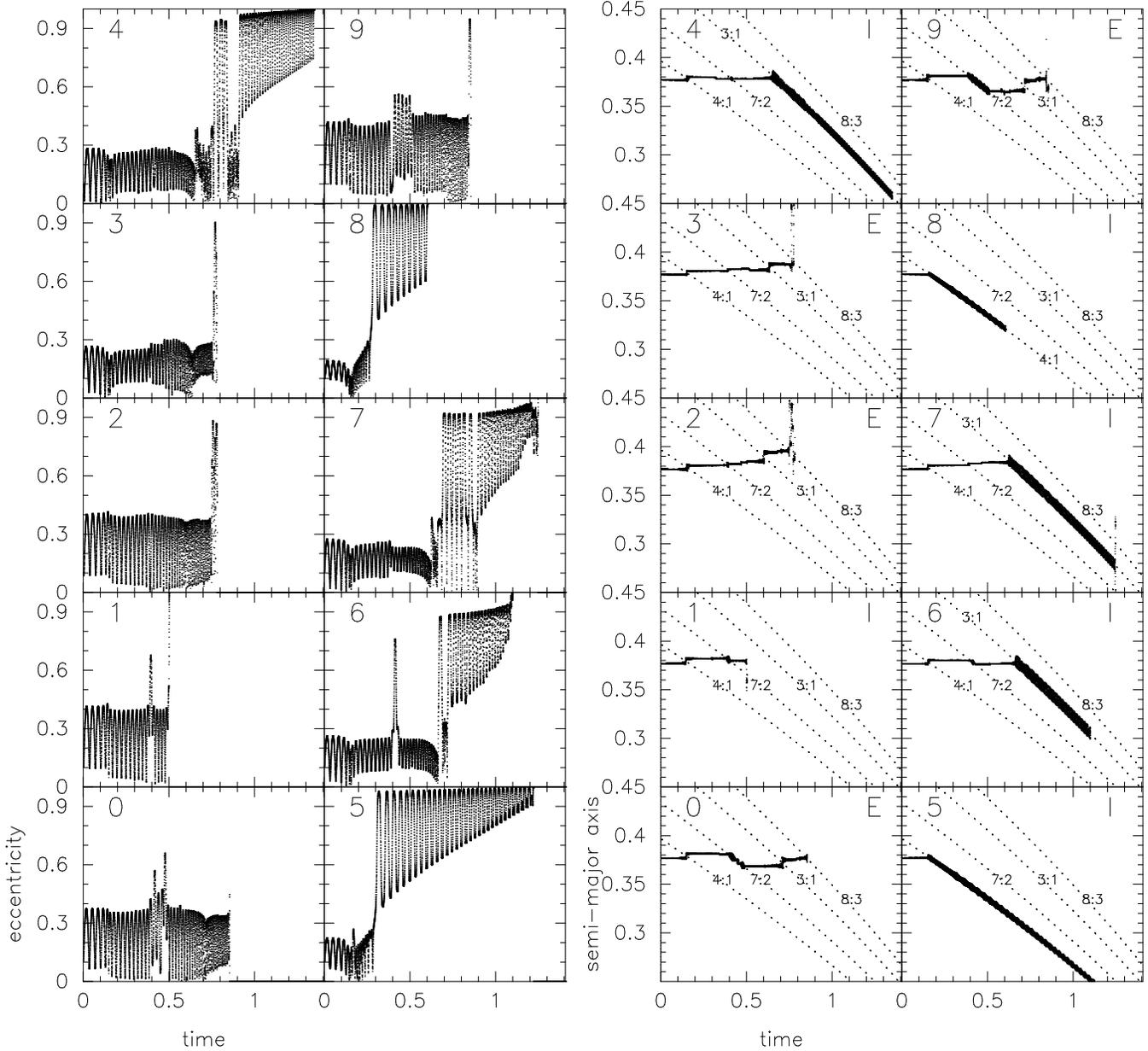

\caption[junk]{
a) Eccentricities as a function of time for 10 particles
as part of the N8 integration (see Tab.~1).  Time is
given in units of $10^5$ periods where a period
corresponds to the initial orbital period of the planet.
The migration rate is $D_a = 3 \times 10^{-6}$, planet eccentricity 
$\epsilon_p = 0.3$, initial particle eccentricity $\epsilon_0 =0.1$, 
and planet mass in units of the stellar mass, $M_p/M_* = 10^{-3}$.
Particles were set initially with
semi-major axes just within the 4:1 resonance.
Particle numbers are labelled on the upper left of each panel.
Particles 5 and 8 spend time trapped in the 4:1 resonance and
impact the star.  Particles 4, 6, and 7 spend time trapped
in the 3:1 resonance and eventually impact the star.  
Prior to impact the surfaces of these particles would be evaporated by 
the star.  Particles 0,2,3, and 9
are ejected when the 8:3 resonance causes them to cross the Hill
sphere of the planet.  Particle 1 impacts the star as a result of
the 10:3 resonance.

b) Particle semi-major axes (in units of the  planet's initial semi-major axis)
as a function of time for the same
10 particles.  
The location of various resonances are shown as
dotted lines and are labelled.  On the upper
right of each box the fate of the particle is shown
where E  refers to ejection by the planet and I refers
to an impact with the star.
While some particles spend time trapped in resonances such
as the 3:1 and 4:1 others are not.  Ejection or impact occurs
during the influence of a resonance.  4 particles
were ejected during this simulation and the remaining 6
impacted the star.
\label{fig:fig2} }
\end{figure*}
\end{document}